\documentclass[a4paper]{amsart}

\RequirePackage{amsmath}
\RequirePackage{bm}
\RequirePackage{amssymb}
\RequirePackage{upref}
\RequirePackage{amsthm}
\RequirePackage{enumerate}
\RequirePackage{pb-diagram}
\RequirePackage{amsfonts}
\RequirePackage[mathscr]{eucal}
\RequirePackage{verbatim}
\RequirePackage{xr}
\RequirePackage{graphicx}
\usepackage{calc}
\usepackage{xspace}
\RequirePackage{color}
\RequirePackage{ifthen}

\newcommand{\cf}{cf.\@\xspace}
\newcommand{\resp}{resp.\@\xspace}

\newcommand{\al}{\alpha}
\newcommand{\bet}{\beta}
\newcommand{\ga}{\gamma}
\newcommand{\de}{\delta }
\newcommand{\e}{\epsilon}

\newcommand{\f}{\varphi}
\newcommand{\h}{\eta}

\newcommand{\ka}{\kappa}

\newcommand{\n}{\nu}
\newcommand{\om}{\omega}

\newcommand{\s}{\sigma}
\newcommand{\x}{\xi}

\newcommand{\D}{\varDelta}
\newcommand{\F}{\varPhi}
\newcommand{\Lam}{\varLambda}
\newcommand{\Om}{\varOmega}

\newcommand{\di}[1]{#1\nobreakdash-\hspace{0pt}dimensional}

\newcommand{\fv}[2]{#1\hspace{0pt}_{|_{#2}}}

\newcommand{\so}{{\mc S_0}}

\newcommand{\const}{\tup{const}}

\newcommand{\msp[1]}[1]{\mspace{#1mu}}

\newcommand{\R}[1][n+1]{{\protect\mathbb R}^{#1}}

\newcommand{\Cc}{{\protect\mathbb C}}
\newcommand{\K}{{\protect\mathbb K}}
\newcommand{\N}{{\protect\mathbb N}}

\newcommand{\eR}{\stackrel{\lower1ex \hbox{\rule{6.5pt}{0.5pt}}}{\msp[3]\R[]}}
\newcommand{\eN}{\stackrel{\lower1ex \hbox{\rule{6.5pt}{0.5pt}}}{\msp[1]\N}}
\newcommand{\eO}{\stackrel{\lower1ex \hbox{\rule{6pt}{0.5pt}}}{\msc O}}

\DeclareMathOperator{\graph}{graph}

\DeclareMathOperator{\supp}{supp}
\DeclareMathOperator{\id}{id}

\newcommand\ra{\rightarrow}

\newcommand\pde[2]{\frac {\partial#1}{\partial#2}}
 
\newcommand\df[2]{\frac {d#1}{d#2}}

\newcommand{\un}{\infty}
\newcommand{\A}{\forall}

\newcommand{\set}[2]{\{\,#1\colon #2\,\}}
\newcommand{\uu}{\cup}
\newcommand{\ii}{\cap}
\newcommand{\uuu}{\bigcup}

\newcommand{\uud}{ \stackrel{\lower 1ex \hbox {.}}{\uu}}
\newcommand{\uuud}[1]{ \stackrel{\lower 1ex \hbox {.}}{\uuu_{#1}}}
\newcommand\su{\subset}
\newcommand\Su{\Subset}

\newcommand{\sminus}[1][28]{\raise 0.#1ex\hbox{$\scriptstyle\setminus$}}

\newcommand{\wed}{\wedge}

\newcommand{\abs}[1]{\lvert#1\rvert}

\newcommand{\norm}[1]{\lVert#1\rVert}

\newcommand{\spd}[2]{\protect\langle #1,#2\protect\rangle}

\newcommand\ch[3]{\varGamma_{#1#2}^#3}
\newcommand\cha[3]{{\bar\varGamma}_{#1#2}^#3}

\newcommand{\riem}[4]{R_{#1#2#3#4}}
\newcommand{\riema}[4]{{\bar R}_{#1#2#3#4}}

\newcommand{\tit}{\textit}

\newcommand{\tup}{\textup}

\newcommand{\mc}{\protect\mathcal}
\newcommand{\msc}{\protect\mathscr}

\providecommand{\bysame}{\makebox[3em]{\hrulefill}\thinspace}

\newcommand{\cq}[1]{\glqq{#1}\grqq\,}

\newcommand{\bt}{\begin{thm}}
\newcommand{\bl}{\begin{lem}}
\newcommand{\bc}{\begin{cor}}
\newcommand{\bd}{\begin{definition}}
\newcommand{\bpp}{\begin{prop}}
\newcommand{\br}{\begin{rem}}
\newcommand{\bn}{\begin{note}}
\newcommand{\be}{\begin{ex}}
\newcommand{\bes}{\begin{exs}}
\newcommand{\bb}{\begin{example}}
\newcommand{\bbs}{\begin{examples}}
\newcommand{\ba}{\begin{axiom}}
\newcommand{\bas}{\begin{assumption}}

\newcommand{\et}{\end{thm}}
\newcommand{\el}{\end{lem}}
\newcommand{\ec}{\end{cor}}
\newcommand{\ed}{\end{definition}}
\newcommand{\epp}{\end{prop}}
\newcommand{\er}{\end{rem}}
\newcommand{\en}{\end{note}}
\newcommand{\ee}{\end{ex}}
\newcommand{\ees}{\end{exs}}
\newcommand{\eb}{\end{example}}
\newcommand{\ebs}{\end{examples}}
\newcommand{\ea}{\end{axiom}}
\newcommand{\eas}{\end{assumption}}

\newcommand{\bp}{\begin{proof}}
\newcommand{\ep}{\end{proof}}
\newcommand{\eps}{\renewcommand{\qed}{}\end{proof}}

\newcommand{\bal}{\begin{align}}

\newcommand{\bi}[1][1.]{\begin{enumerate}[\upshape #1]}
\newcommand{\bia}[1][(1)]{\begin{enumerate}[\upshape #1]}
\newcommand{\bin}[1][1]{\begin{enumerate}[\upshape\bfseries #1]}
\newcommand{\bir}[1][(i)]{\begin{enumerate}[\upshape #1]}
\newcommand{\bic}[1][(i)]{\begin{enumerate}[\upshape\hspace{2\cma}#1]}
\newcommand{\bis}[2][1.]{\begin{enumerate}[\upshape\hspace{#2\parindent}#1]}
\newcommand{\ei}{\end{enumerate}}

\newcommand\ndots{\raise 0.47ex \hbox {,}\hskip0.06em\cdots %
     \raise 0.47ex \hbox {,}\hskip0.06em} 

\newcommand{\q}{\quad}
\newcommand{\qq}{\qquad}

\newcommand\nd{\noindent}

\newskip\Csmallskipamount                                                
\Csmallskipamount=\smallskipamount
\newskip\Cmedskipamount
\Cmedskipamount=\medskipamount
\newskip\Cbigskipamount
\Cbigskipamount=\bigskipamount

\newcommand\cvs{\vspace\Csmallskipamount}   
\newcommand\cvm{\vspace\Cmedskipamount}

\newskip\csa
\csa=\smallskipamount

\newskip\cma
\cma=\medskipamount

\newskip\cba
\cba=\bigskipamount

\newdimen\spt
\newcommand\citem{\cvs\advance\itemno by
1{(\romannumeral\the\itemno})\hskip3pt}
\newcommand{\bitem}{\cvm\nd\advance\itemno by
1{\bf\the\itemno}\hspace{\cma}}

\newcount\itemno
\newcommand{\las}[1]{\label{S:#1}}

\newcommand{\lae}[1]{\label{E:#1}}
\newcommand{\lat}[1]{\label{T:#1}}
\newcommand{\lal}[1]{\label{L:#1}}

\newcommand{\rs}[1]{Section~\ref{S:#1}}

\newcommand{\rt}[1]{Theorem~\ref{T:#1}}
\newcommand{\rl}[1]{Lemma~\ref{L:#1}}

\newcommand{\re}[1]{\eqref{E:#1}}
\newcommand{\frt}[1]{Theorem~\ref{T:#1} on page~\tup{\pageref{T:#1}}}

\newcommand{\fre}[1]{\eqref{E:#1} on page~\tup{\pageref{E:#1}}}

\newskip\thmskip
\thmskip=\parindent

\newskip\hsk
\newenvironment{hinw}{\labelsep=0pt\begin{list}{}{\labelsep=0pt\itemindent=0pt\labelwidth=0pt\leftmargin=\parindent\rightmargin=0pt\partopsep=\cba}%
\item\it\nopagebreak\nopagebreak}%
{\end{list}}

\newcommand\bh{\begin{hinw}}
\newcommand{\eh}{\end{hinw}}

\newtheoremstyle{normal}
  {\cba}
  {\cba}
  {}
  {\thmskip}
  {\bfseries}
  {.}
  {\hsk}
  {}

\newtheoremstyle{abschnitt}
  {\cba}
  {\cba}
  {}
  {\thmskip}
  {\bfseries}
  {.}
  {\hsk}
  {}

\newtheoremstyle{italic}
  {\cba}
  {\cba}
  {\itshape}
  {\thmskip}
  {\bfseries}
  {.}
  {\hsk}
  {}

\newtheoremstyle{aufgaben}
  {\cba}
  {\cba}
  {}
  {}
  {\normalsize\bfseries}
  {.}
  {\hsk}
  {}

\newtheoremstyle{break}
  {\cba}
  {\cba}
  {\itshape}
  {}
  {\bfseries}
  {.}
  {\newline}
  {}

\swapnumbers
\theoremstyle{italic}
\newtheorem{thm}[subsection]{Theorem}
\newtheorem{lem}[subsection]{Lemma}
\newtheorem{prop}[subsection]{Proposition}
\newtheorem{cor}[subsection]{Corollary}

\theoremstyle{normal}
\newtheorem{rem}[subsection]{Remark}
\newtheorem{definition}[subsection]{Definition}
\newtheorem{example}[subsection]{Example}
\newtheorem{examples}[subsection]{Examples}
\newtheorem{ex}[subsection]{Exercise}
\newtheorem{note}[subsection]{}
\newtheorem{axiom}[subsection]{Axiom}
\newtheorem{assumption}[subsection]{Assumption}

\theoremstyle{aufgaben}
\newtheorem{exs}[subsection]{Exercises}

\swapnumbers

\numberwithin{equation}{section}
\numberwithin{figure}{section}

\newenvironment{textequation}[1][0.8]
{\begin{equation}
\begin{aligned}
\begin{minipage}{#1\linewidth}}
{\end{minipage}
\end{aligned}
\end{equation}
\ignorespacesafterend}

\newcommand{\btext}{\begin{textequation}}
\newcommand{\etext}{\end{textequation}}

\def\hinweis{\@startsection{subsection}{2}%
 \z@{0.7\linespacing\@plus 0.5\linespacing}{0.7\linespacing}%
{\normalfont\itshape\indent}}

\newcounter{hours}\newcounter{minutes}
\newcommand{\printtime}{%
\setcounter{hours}{\time/60}%
\setcounter{minutes}{\time-\value{hours}*60}%
\ifthenelse{\value{minutes}<10}{\thehours :0\theminutes}{\thehours:\theminutes}}

\usepackage[german,english]{babel}
\usepackage{graphicx}
\RequirePackage{amsmath}
\RequirePackage{bm}
\RequirePackage{amssymb}
\RequirePackage{upref}
\RequirePackage{amsthm}
\RequirePackage{enumerate}
\RequirePackage{pb-diagram}
\RequirePackage{amsfonts}
\RequirePackage[mathscr]{eucal}
\RequirePackage{verbatim}
\RequirePackage{xr}
\RequirePackage{graphicx}
\usepackage{calc}
\usepackage{xspace}

\makeatletter
\RequirePackage{color}
\newcommand{\ann}[1]{\renewcommand{\@makefnmark}{\mbox{$^{\color{red}{\@thefnmark}}$}}%
\footnote {#1}}
\makeatother

\RequirePackage{upref}
\RequirePackage{amsthm}
\RequirePackage{enumerate}
\usepackage[mathscr]{eucal}

\usepackage{xr-hyper}

\usepackage{calc}

\newlength{\oddsidemarginlength}
\newlength{\topmarginlength}

\newcounter{numberoflines}
\newcounter{tempcc}
\oddsidemargin=\oddsidemarginlength
\evensidemargin=\oddsidemargin
\topmargin=\topmarginlength
\usepackage[colorlinks=true,linkcolor=blue,citecolor=blue,urlcolor=blue]{hyperref}  

\begin{document}

\flushbottom


\title{The quantization of gravity in globally hyperbolic spacetimes}

\author{Claus Gerhardt}
\address{Ruprecht-Karls-Universit\"at, Institut f\"ur Angewandte Mathematik,
Im Neuenheimer Feld 294, 69120 Heidelberg, Germany}
\email{\href{mailto:gerhardt@math.uni-heidelberg.de}{gerhardt@math.uni-heidelberg.de}}
\urladdr{\href{http://www.math.uni-heidelberg.de/studinfo/gerhardt/}{http://www.math.uni-heidelberg.de/studinfo/gerhardt/}}

%
\subjclass[2000]{83,83C,83C45}
\keywords{globally hyperbolic Lorentzian manifold, quantum gravity, unification, unified quantum theory}
\date{\today}
%


\begin{abstract} 
We apply the ADM approach to obtain a Hamiltonian description of the Einstein-Hilbert action. In doing so we add four new ingredients: (i) We eliminate the diffeomorphism constraints. (ii) We replace the densities $\sqrt g$ by a function $\f(x,g_{ij})$ with the help of a fixed metric $\chi$ such that the Lagrangian and hence the Hamiltonian are functions. (iii) We consider the Lagrangian to be defined in a fiber bundle with base space $\so$ and fibers F(x) which can be treated as Lorentzian manifolds equipped with the Wheeler-DeWitt metric. It turns out that the fibers are globally hyperbolic. (iv) The Hamiltonian operator $H$ is a normally hyperbolic operator in the bundle acting only in the fibers and the Wheeler-DeWitt equation $Hu=0$ is a hyperbolic equation in the bundle. Since the corresponding Cauchy problem can be solved for arbitrary smooth data with compact support, we then  apply the standard techniques of QFT which can be naturally modified to work in the bundle.
\end{abstract}

\maketitle

\tableofcontents

\setcounter{section}{0}
\section{Introduction}
QFT has been very successful in quantizing non-gravitational fields while ignoring the interaction with gravity. In the attempts to quantize gravity, on the other hand, canonical quantization has mostly been utilized which requires to switch from the Lagrangian to the Hamiltonian viewpoint. However, the Lagrangian is degenerate resulting in two constraints, the diffeomorphism constraint and the Hamiltonian constraint. The diffeomorphism constraint is usually ignored since nobody knows how to handle it. The Hamiltonian constraint leads to the Wheeler-DeWitt equation which could only be solved by assuming a high degree of symmetry.

In this paper we use canonical quantization to obtain a setting in which the standard techniques of QFT can be applied to achieve quantization of the gravitational field, i.e., gravity can be treated like a non-gravitational field.

In order to make this approach work four new ideas had to be introduced in the process of canonical quantization:

\cvm
(i) We eliminated the diffeomorphism constraint by proving that it suffices to consider metrics that split according to
\begin{equation}
d\bar s^2=-w^2 (dx^0)^2+g_{ij}dx^idx^j
\end{equation}
after introducing a global time function $x^0$. The underlying spacetime $N=N^{n+1}$ can be considered to be a topological product
\begin{equation}
N=I\times\so
\end{equation}
where $I\su\R[]$ is an open interval, $\so$ a Cauchy hypersurface, fixed for all metrics under consideration, and $g_{ij}=g_{ij}(x^0,x)$, $x\in\so$, a Riemannian metric.

\cvm
(ii) The volume element $\sqrt g$, $g=\det(g_{ij})$, is a density and it appears explicitly in the Lagrangian and in the Hamiltonian. However, the Hamiltonian has to be an invariant, i.e., a function and not a density. To overcome this difficulty we fixed a metric $\chi\in T^{0,2}(\so)$ and defined the function $\f$ by
\begin{equation}
\f^2=\frac{\det(g_{ij})}{\det(\chi_{ij})}
\end{equation}
such that $\f=\f(x,g_{ij})$ and
\begin{equation}
\sqrt g=\f\sqrt\chi.
\end{equation}
The density $\sqrt\chi$ will be later ignored when performing the Legendre transformation in accordance with Mackey's advice to only use rectangular coordinates in canonical quantization, \cf \cite[p. 94]{mackey:book}.

\cvm
(iii) After the Legendre transformation the momenta depend on $x\in\so$. To overcome this difficulty we consider a fiber bundle with base space $\so$ where the fibers are the positive definite metrics $g_{ij}(x)$ over $x$, i.e., a fiber $F(x)$ is an open, convex cone in a finite dimensional vector space. We treat this cone as a manifold endowing it with the DeWitt metric which is Lorentzian. It turns out that $F(x)$ is globally hyperbolic. Let us call the bundle $E$. Each fiber has a Cauchy hypersurface $M(x)$ and we denote the corresponding bundle  by $\hat E$.

DeWitt \cite{dewitt:qg} had already the idea to consider $F(x)$ as a Lorentzian space but he did not consider it to be a fiber and his DeWitt metric was no real tensor since it contained a density.

The introduction of the bundle $E$ simplifies the mathematical model after canonical quantization dramatically. The Hamiltonian operator $H$ is a normally hyperbolic differential operator acting only in the fibers which are globally hyperbolic spacetimes and the Wheeler-DeWitt equation is the hyperbolic equation
\begin{equation}
Hu=0,
\end{equation}
where $u$ is defined in $E$.

The Cauchy problem
\begin{equation}\lae{1.6}
\begin{aligned}
Hu&=f\\
\fv uM&=u_0\\
\fv{D_\nu u}M&=u_1
\end{aligned}
\end{equation}
is uniquely solvable in $E$ with $u\in C^\un(E,\K)$, $\K=\R[]\vee\K=\Cc$, for arbitrary
\begin{equation}
u_0,u_1\in C^\un_c(\hat E,\K)\q\wed\q f\in C^\un_c(E,\K).
\end{equation}

\cvm
(iv) In view of \re{1.6} the standard techniques of QFT, slightly modified to accept the present setting, can be applied to construct a quantum field $\F_{\hat E}$ which maps functions $u\in C^\un_c(E,\R[])$ to self-adjoint operators in the symmetric Fock space created from the Hilbert space
\begin{equation}
H_{\hat E}=L^2(\hat E,\Cc).
\end{equation}
The quantum field also satisfies the Wheeler-DeWitt equation in the distributional sense. 

This was a summary of the contents of Sections 3-6. In the last section we consider the interaction of gravity with a scalar field $y$
\begin{equation}
y:N\ra M_1,
\end{equation}
where $M_1$ is a complete Riemannian space. The corresponding Lagrangian is
\begin{equation}
L_1=-\tfrac12 \bar g^{\al\bet}y^A_\al y^B_\bet g_{AB} -V(y).
\end{equation}
It turns out that the combined fields can be treated just like gravity alone. The fibers have to be replaced by
\begin{equation}
F\times M_1
\end{equation}
which are again globally hyperbolic, the Hamiltonian is a normally hyperbolic differential operator and the further reasoning is identical to the former one without a scalar field.
\br
The fibers $F$, and also $F\times M_1$, are spacetimes with a past crushing singularity, a big bang, \cf \frt{4.3}.
\er
\section{Definitions and notations}
The main objective of this section is to state the equations of Gau{\ss}, Codazzi,
and Weingarten for spacelike hypersurfaces $M$ in a \di {(n+1)} Lorentzian
manifold
$N$.  Geometric quantities in $N$ will be denoted by
$(\bar g_{ \al \bet}),(\riema  \al \bet \ga \de)$, etc., and those in $M$ by $(g_{ij}), 
(\riem ijkl)$, etc.. Greek indices range from $0$ to $n$ and Latin from $1$ to $n$;
the summation convention is always used. Generic coordinate systems in $N$ resp.
$M$ will be denoted by $(x^ \al)$ \resp $(\x^i)$. Covariant differentiation will
simply be indicated by indices, only in case of possible ambiguity they will be
preceded by a semicolon, i.e., for a function $u$ in $N$, $(u_ \al)$ will be the
gradient and
$(u_{ \al \bet})$ the Hessian, but e.g., the covariant derivative of the curvature
tensor will be abbreviated by $\riema  \al \bet \ga{ \de;\e}$. We also point out that
\begin{equation}
\riema  \al \bet \ga{ \de;i}=\riema  \al \bet \ga{ \de;\e}x_i^\e
\end{equation}
with obvious generalizations to other quantities.

Let $M$ be a \tit{spacelike} hypersurface, i.e., the induced metric is Riemannian,
with a differentiable normal $\n$ which is timelike.

In local coordinates, $(x^ \al)$ and $(\x^i)$, the geometric quantities of the
spacelike hypersurface $M$ are connected through the following equations
\begin{equation}\lae{01.2}
x_{ij}^ \al= h_{ij}\n^ \al
\end{equation}
the so-called \tit{Gau{\ss} formula}. Here, and also in the sequel, a covariant
derivative is always a \tit{full} tensor, i.e.

\begin{equation}
x_{ij}^ \al=x_{,ij}^ \al-\ch ijk x_k^ \al+ \cha  \bet \ga \al x_i^ \bet x_j^ \ga.
\end{equation}
The comma indicates ordinary partial derivatives.

In this implicit definition the \tit{second fundamental form} $(h_{ij})$ is taken
with respect to $\n$.

The second equation is the \tit{Weingarten equation}
\begin{equation}
\n_i^ \al=h_i^k x_k^ \al,
\end{equation}
where we remember that $\n_i^ \al$ is a full tensor.

Finally, we have the \tit{Codazzi equation}
\begin{equation}
h_{ij;k}-h_{ik;j}=\riema \al \bet \ga \de\n^ \al x_i^ \bet x_j^ \ga x_k^ \de
\end{equation}
and the \tit{Gau{\ss} equation}
\begin{equation}
\riem ijkl=- \{h_{ik}h_{jl}-h_{il}h_{jk}\} + \riema  \al \bet\ga \de x_i^ \al x_j^ \bet
x_k^ \ga x_l^ \de.
\end{equation}

Now, let us assume that $N$ is a globally hyperbolic Lorentzian manifold with a
 Cauchy surface. 
$N$ is then a topological product $I\times \mc S_0$, where $I$ is an open interval,
$\mc S_0$ is a  Riemannian manifold, and there exists a Gaussian coordinate
system
$(x^ \al)$, such that the metric in $N$ has the form 
\begin{equation}\lae{01.7}
d\bar s_N^2=e^{2\psi}\{-{dx^0}^2+\s_{ij}(x^0,x)dx^idx^j\},
\end{equation}
where $\s_{ij}$ is a Riemannian metric, $\psi$ a function on $N$, and $x$ an
abbreviation for the spacelike components $(x^i)$. 
We also assume that
the coordinate system is \tit{future oriented}, i.e., the time coordinate $x^0$
increases on future directed curves. Hence, the \tit{contravariant} timelike
vector $(\x^ \al)=(1,0,\dotsc,0)$ is future directed as is its \tit{covariant} version
$(\x_ \al)=e^{2\psi}(-1,0,\dotsc,0)$.

Let $M=\graph \fv u\so$ be a spacelike hypersurface
\begin{equation}
M=\set{(x^0,x)}{x^0=u(x),\,x\in\mc S_0},
\end{equation}
then the induced metric has the form
\begin{equation}
g_{ij}=e^{2\psi}\{-u_iu_j+\s_{ij}\}
\end{equation}
where $\s_{ij}$ is evaluated at $(u,x)$, and its inverse $(g^{ij})=(g_{ij})^{-1}$ can
be expressed as
\begin{equation}\lae{01.10}
g^{ij}=e^{-2\psi}\{\s^{ij}+\frac{u^i}{v}\frac{u^j}{v}\},
\end{equation}
where $(\s^{ij})=(\s_{ij})^{-1}$ and
\begin{equation}\lae{01.11}
\begin{aligned}
u^i&=\s^{ij}u_j\\
v^2&=1-\s^{ij}u_iu_j\equiv 1-\abs{Du}^2.
\end{aligned}
\end{equation}
Hence, $\graph u$ is spacelike if and only if $\abs{Du}<1$.

The covariant form of a normal vector of a graph looks like
\begin{equation}
(\n_ \al)=\pm v^{-1}e^{\psi}(1, -u_i).
\end{equation}
and the contravariant version is
\begin{equation}
(\n^ \al)=\mp v^{-1}e^{-\psi}(1, u^i).
\end{equation}
Thus, we have
\br Let $M$ be spacelike graph in a future oriented coordinate system. Then the
contravariant future directed normal vector has the form
\begin{equation}
(\n^ \al)=v^{-1}e^{-\psi}(1, u^i)
\end{equation}
and the past directed
\begin{equation}\lae{01.15}
(\n^ \al)=-v^{-1}e^{-\psi}(1, u^i).
\end{equation}
\er

In the Gau{\ss} formula \re{01.2} we are free to choose the future or past directed
normal, but we stipulate that we always use the past directed normal.
Look at the component $ \al=0$ in \re{01.2} and obtain in view of \re{01.15}

\begin{equation}\lae{01.16}
e^{-\psi}v^{-1}h_{ij}=-u_{ij}- \cha 000\mspace{1mu}u_iu_j- \cha 0j0
\mspace{1mu}u_i- \cha 0i0\mspace{1mu}u_j- \cha ij0.
\end{equation}
Here, the covariant derivatives are taken with respect to the induced metric of
$M$, and
\begin{equation}
-\cha ij0=e^{-\psi}\bar h_{ij},
\end{equation}
where $(\bar h_{ij})$ is the second fundamental form of the hypersurfaces
$\{x^0=\const\}$.

An easy calculation shows
\begin{equation}
\bar h_{ij}e^{-\psi}=-\tfrac{1}{2}\dot\s_{ij} -\dot\psi\s_{ij},
\end{equation}
where the dot indicates differentiation with respect to $x^0$.

\section{The Wheeler-DeWitt equation}\las{3}

Let $N=N^{n+1}$ be a globally hyperbolic spacetime. We consider the functional
\begin{equation}
J=\int_N(\bar R-2\Lam),
\end{equation}
where $\bar R$ is the scalar curvature and $\Lam$ a cosmological constant. The  integration over $N$ is to be understood only symbolically since we are only interested in the first variation of the functional, i.e., when a metric $\bar g=(\bar g_{\al\bet})$ in $N$ is given, we are only interested in the first variation of $J$ with respect to \tit{compact} variations of $\bar g$, hence it suffices to integrate only over open and precompact subsets $\Om\su N$ such that
\begin{equation}
J=\int_\Om(\bar R-2\Lam).
\end{equation}

It is well known that, when the first variation of $J$ with respect to arbitrary compact variations of $\bar g$ vanishes, the metric $\bar g$ satisfies the Einstein equations with cosmological constant $\Lam$, namely,
\begin{equation}
G_{\al\bet}+\Lam \bar g_{\al\bet}=0,
\end{equation}
where $G_{\al\bet}$ is the Einstein tensor.

When $N$ endowed with a metric $\bar g$ is globally hyperbolic, there exists a global time function $f\in C^\un(N)$ such that
\begin{equation}
\norm{Df}^2=\bar g^{\al\bet}f_\al f_\bet<0,
\end{equation}
$N$ can be written as a topological product
\begin{equation}
N=I\times \so,\q I=(a,b)\su\R[],
\end{equation}
where
\begin{equation}
\so=f^{-1}(c),\q a<c<b,
\end{equation}
is a Cauchy hypersurface and there exists a Gaussian coordinate system $(x^\al)$, $0\le\al\le n$, such that $x^0=f$ and the metric $\bar g$ splits according to
\begin{equation}\lae{3.7}
d\bar s^2=-w^2(dx^0)^2+\bar g_{ij}dx^idx^j,
\end{equation}
where $(x^i)$, $1\le i\le n$, are local coordinates of $\so$ and
\begin{equation}
\bar g_{ij}=\bar g_{ij}(x^0,x),\q x\in\so,
\end{equation}
are Riemannian and $w>0$ is function.

Without loss of generality we may always assume that $0\in I$ and that $c=0$. When there exists a time function and an associated Gaussian coordinate system such that \re{3.7} is valid we also say that $x^0$ splits the metric.

\bl
Let $(N,\bar g)$ be a globally hyperbolic spacetime and let $f$ be a time function that splits $\bar g$. Let $\om=(\om_{\al\bet})$ be an arbitrary smooth symmetric tensor field with compact support and define
\begin{equation}
\bar g(\e)=\bar g+\e\,\om
\end{equation}
for small values of $\e$
\begin{equation}
\abs{\e}<\e_0.
\end{equation}
If $\e_0$ is small enough, the tensor fields $\bar g(\e)$ will also be Lorentzian metrics that will be split by $f$.
\el
\bp
We shall only prove that the $\bar g(\e)$ will be split by $f$, since the other claim is obvious.

Define the conformal \tit{covariant} metrics
\begin{equation}
g(\e)=\abs{\spd{Df}{Df}}^{2}\bar g(\e),
\end{equation}
then $Df$ is a unit gradient field for each $g(\e)$. Let $\so=f^{-1}(0)$ and consider the flow $x=x(t,\xi)$ satisfying\begin{equation}
\begin{aligned}
\dot x&=-Df,\\
x(0,\xi)&=\xi,
\end{aligned}
\end{equation}
where $\xi\in \so$. For fixed $\xi$ the flow is defined on a maximal time interval $J=(a_0,b_0)$. If we can prove that $J=I=(a,b)=f(N)$, then we would have proved that each metric $g(\e)$ satisfies
\begin{equation}
d\bar s^2=-(dx^0)^2+g_{ij}dx^dx^j,
\end{equation}
where the $g_{ij}$ are Riemannian and depend smoothly on $\e$, \cf the arguments in \cite[p. 27]{cg:cp}. 

It suffices to prove $b_0=b$. Assume that
\begin{equation}
b_0<b,
\end{equation}
and let $K$ be the support of $\om$. Then there exists $t_0<b_0$ such that
\begin{equation}
x(t,\xi)\notin K\qq\A\,t>t_0,
\end{equation}
for otherwise there would exist a sequence $(t_k)$
\begin{equation}
t_k\ra b_0\q\wed\q x(t_k,\xi)\in K
\end{equation}
contradicting the maximality of $J$, since there has to be a \cq{singularity} for the flow in $b_0$.

Thus, choose
\begin{equation}
t_0<t_1<b_0,
\end{equation}
then
\begin{equation}
x(t_1,\xi)\in M(t_1)=f^{-1}(t_1),
\end{equation}
because
\begin{equation}
f(x(t,\xi))=t
\end{equation}
as one easily checks.

Let $y=y(t,\xi)$ be the flow corresponding to $\e=0$, then $y$ covers $N$ by assumption and hence there exists $\zeta\in\so$ such that
\begin{equation}
y(t_1,\zeta)=x(t_1,\xi).
\end{equation}
Then the integral curve
\begin{equation}
\begin{aligned}
\dot y&=-Df,\\
y(t_1,\zeta)&=x(t_1,\xi),
\end{aligned}
\end{equation}
where the contravariant vector is now defined with the help $\bar g=\bar g(0)$
\begin{equation}
D^\al f=\bar g^{\al\bet}f_\bet,
\end{equation}
would be a smooth continuation of $x(t,\xi)$ past $b_0$, a contradiction.
\ep

The preceding lemma will enable us to eliminate the so-called diffeomorphism constraint when switching from a Lagrangian to a Hamiltonian view of gravity.

\bt\lat{3.2}
Let $(N,\bar g)$ be a globally hyperbolic spacetime, $f$ a time function that splits $\bar g$ with Cauchy hypersurface $\so$. Let $\tilde \Om\Su N$ be open and precompact and assume that the first variation of the functional
\begin{equation}
J=\int_{\tilde\Om} (\bar R-2\Lam)
\end{equation}
vanishes in $\bar g$ for those compact variations of $\bar g$ which can be expressed in the form
\begin{equation}\lae{3.21}
d\bar s^2=-w^2(dx^0)^2+g_{ij}dx^idx^j,
\end{equation}
where $(g_{ij}(x^0,x))$ is Riemannian, then the first variation of $J$ in $\bar g$ also vanishes for arbitrary compact variations.
\et
\bp
Let $\om=(\om_{\al\bet})$ be an arbitrary smooth symmetric tensor with compact support in $\tilde\Om$. The metrics
\begin{equation}
\bar g(\e)=\bar g+\e\,\om,\qq\abs \e<\e_0,
\end{equation}
then satisfy \re{3.21} for small $\e_0$, in view of the preceding lemma, hence
\begin{equation}
\de J(\bar g;\dot g(0))=0.
\end{equation}
But the first variation is a scalar, hence
\begin{equation}
\de J(\bar g;\dot g(0))=\de J(\bar g;\om),
\end{equation}
where at the right-hand side we used an arbitrary coordinate system to express the tensors.
\ep 

We are now ready to look at the Hamiltonian form of the Einstein-Hilbert action following \cite{adm:old}.

Let $\tilde\Om\Su N$ be an arbitrary open, precompact set. Then we consider the functional
\begin{equation}
J=\al_N^{-1}\int_{\tilde\Om} (\bar R-2\Lam),
\end{equation}
where $\al_N$ is a positive constant and we assume that there exists a time function $f=x^0$ in $N$ with Cauchy hypersurface $\so=f^{-1}(0)$ and where, in view of \rt{3.2}, we only consider metrics of the form
\begin{equation}\lae{3.26}
d\bar s^2=-w^2(dx^0)^2+ g_{ij}dx^idx^j,
\end{equation}
where $w$ is an arbitrary smooth positive function and $g_{ij}=g_{ij}(x^0,x)$, $x\in \so$, Riemannian metrics. Let us fix a metric $\bar g=(\bar g_{\al\bet})$ as in \re{3.26}, then we deduce from the Gau{\ss} equation
\begin{equation}\lae{3.27}
\bar R=H^2-\abs{A}^2+R-2\bar R_{\al\bet}\nu^\al\nu^\bet,
\end{equation}
where $R$ is the scalar curvature of the slices
\begin{equation}
M(t)=\{x^0=t\},
\end{equation}
$H$ the mean curvature of $M(t)$
\begin{equation}\lae{3.29}
H=g^{ij}h_{ij}=\sum_{i=1}^n\ka_i,
\end{equation}
where $\ka_i$ are the principal curvatures, $\abs A^2$ is defined by
\begin{equation}\lae{3.30}
\abs A^2=h_{ij}h^{ij}=\sum_{i=1}^n\ka_i^2,
\end{equation}
and where the second fundamental form $h_{ij}$ of $M(t)$ can be expressed as
\begin{equation}\lae{3.31}
h_{ij}=-\tfrac 12 \dot g_{ij}w^{-1},
\end{equation}where
\begin{equation}
\dot g_{ij}=\pde {g_{ij}}t,
\end{equation}
when we identify $t$ with $x^0$.

The last term on the right-hand side of \re{3.27}  can be written as
\begin{equation}
-2\bar R_{\al\bet}\nu^\al\nu^\bet=-2(H^2-\abs A^2)+D_\al a^\al,
\end{equation}
\cf \cite[equ. (4.60)]{kiefer:book}. Since the divergence term can be neglected the functional $J$ is equal to
\begin{equation}
J=\al_N^{-1}\int_a^b\int_\Om\{\abs A^2-H^2+R-2\Lam\}w\sqrt g,
\end{equation}
where we may assume that
\begin{equation}
\tilde \Om=(a,b)\times\Om,\q\Om\Su\so,
\end{equation}
and
\begin{equation}
(a,b)\Su x^0(N)=I.
\end{equation}

Using the relations \re{3.29}--\re{3.31} we finally conclude 
\begin{equation}\lae{3.37}
\begin{aligned}
J=\al_N^{-1}\int_a^b\int_\Om\{\tfrac14 G^{ij,kl}\dot g_{ij}\dot g_{kl} w^{-2}+R-2\Lam\}w\sqrt g,
\end{aligned}
\end{equation}
where
\begin{equation}\lae{3.38}
G^{ij,kl}=\tfrac12\{g^{ik}g^{jl}+g^{il}g^{jk}\}-g^{ij}g^{kl}
\end{equation}
and
\begin{equation}
(g^{ij})=(g_{ij})^{-1}.
\end{equation}
The metric in \re{3.38} is known as the DeWitt metric and it is a Lorentzian metric as one can easily check.

We are almost ready to define the corresponding Hamiltonian, we only need two adjustments. First, we like to replace the density
\begin{equation}
g=\det(g_{ij})
\end{equation}
by
\begin{equation}
g=\frac{\det(g_{ij})}{\det\chi_{ij})}\det(\chi_{ij})\equiv \f^2\det(\chi_{ij}),
\end{equation}
where $\chi=(\chi_{ij})$ is an arbitrary but fixed Riemannian metric in $\so$ and where $\f$ is now a function
\begin{equation}\lae{3.42}
0<\f=\f(x,g_{ij})=\frac{\sqrt g}{\sqrt\chi}.
\end{equation}

Second, the Riemannian metrics $g_{ij}(t,\cdot)$ are elements of the bundle $T^{0,2}(\so)$. Denote by $E$ the fiber bundle with base $\so$ where the fibers consists of the Riemannian metrics $(g_{ij})$. We shall consider each fiber to be a Lorentzian manifold equipped with the DeWitt metric. Each fiber $F$ has dimension
\begin{equation}
\dim F=\frac{n(n+1)}2\equiv m+1.
\end{equation}
Let $(\xi^a)$, $0\le a\le m$, be  coordinates for a local trivialization such that
\begin{equation}
g_{ij}(x,\xi^a)
\end{equation}
is a local embedding. The DeWitt metric is then expressed as
\begin{equation}
G_{ab}=G^{ij,kl}g_{ij,a}g_{kl,b},
\end{equation}
where a comma indicates partial differentiation.  
In the new coordinate system the curves 
\begin{equation}
t\ra g_{ij}(t,x)
\end{equation}
can be written in the form
\begin{equation}
t\ra \xi^a(t,x)
\end{equation}
and we infer
\begin{equation}
G^{ij,kl}\dot g_{ij}\dot g_{kl}=G_{ab}\dot\xi^a\dot\xi^b. 
\end{equation}
Hence, we can express \re{3.37} as
\begin{equation}\lae{3.49}
J=\int_a^b\int_\Om \al_n^{-1}\{\tfrac14 G_{ab}\dot\xi^a\dot\xi^bw^{-1}\f+(R-2\Lam)w\f\},
\end{equation}
where we now refrain from writing down the density $\sqrt\chi$ explicitly, since it does not depend on $(g_{ij})$ and therefore should not be part of the Legendre transformation.  Here we follow Mackey's advice in \cite[p. 94]{mackey:book} to always consider rectangular coordinates when applying canonical quantization, which can be rephrased that the Hamiltonian has to be a coordinate invariant, hence no densities are allowed.

Denoting the Lagrangian \tit{function} in \re{3.49} by $L$, we define
\begin{equation}
\pi_a= \pde L{\dot\xi^a}=\f G_{ab}\frac1{2\al_N}\dot\xi^bw^{-1}
\end{equation}
and we obtain for the Hamiltonian function $\hat H$
\begin{equation}
\begin{aligned}
\hat H&=\dot\xi^a\pde L{\dot\xi^a}-L\\
&=\f G_{ab}\big(\frac1{2\al_N}\dot\xi^aw^{-1}\big)\big(\frac1{2\al_N}\dot\xi^bw^{-1}\big) w\al_N-\al_N^{-1}(R-2\Lam)\f w\\
&=\f^{-1}G^{ab}\pi_a\pi_b w\al_N-\al^{-1}_N(R-2\Lam)\f w\\
&\equiv H w,
\end{aligned}
\end{equation}
where $G^{ab}$ is the inverse metric. Since $w$ is an arbitrary function we obtain the Hamiltonian constraint
\begin{equation}\lae{3.52}
H=\al_N\f^{-1}G^{ab}\pi_a\pi_b -\al^{-1}_N(R-2\Lam)\f =0.
\end{equation}
Applying canonical quantization, by setting $\hbar=1$, we replace
\begin{equation}
\pi_a=\pi_a(x)\ra \frac1i\pde{}{\xi^a(x)},
\end{equation}
where $(\xi^a(x))$ are points in the fiber over $x\in\so$ and
\begin{equation}
\pde{}{\xi^a(x)}
\end{equation}
denotes partial differentiation in the fiber over $x$.

Each fiber can be viewed as a Lorentzian manifold equipped with the metric
\begin{equation}
\al_N^{-1}\f G_{ab}.
\end{equation}
After quantization the Hamiltonian function is transformed into the hyperbolic differential operator
\begin{equation}\lae{3.56}
\begin{aligned}
H&=-\D-\al^{-1}_N(R-2\Lam)\f\\
&\equiv\square -\al^{-1}_N(R-2\Lam)\f
\end{aligned}
\end{equation}
where $\square$ is the d'Alembertian operator for that metric.

We want to emphasize that $H$ is defined in the bundle $E$ acting only in the fibers. Hence, let $u$ be a smooth function defined in $E$
\begin{equation}
u=u(x,\xi(x))=u(x,g_{ij}(x)),
\end{equation}
then the Hamiltonian condition \re{3.52} takes the form
\begin{equation}
\square u-\al^{-1}_N(R-2\Lam)\f u=0.
\end{equation}
Since each fiber is equipped with the Lorentz metric $(\al^{-1}_N\f G_{ab})$ there is a natural volume element
\begin{equation}\lae{3.59}
\sqrt{\abs{\det(\al^{-1}_N\f G_{ab})}} d\xi
\end{equation}
and for functions $u,v\in  C^\un_c(E,\K)$, where 
\begin{equation}
\K=\Cc\q\vee\q\K=\R[],
\end{equation}
we can define the scalar product
\begin{equation}
\spd uv_E=\int_\so\int_{F(x)}u\bar v,
\end{equation}
where the volume element of the fiber $F(x)$ is given in \re{3.59} and that of $\so$ is defined by the metric $\chi_{ij}$
\begin{equation}
\sqrt\chi dx.
\end{equation}
It is immediately clear that the hyperbolic differential operator is formally self-adjoint, and since each fiber is globally hyperbolic, which will be proved in \rs{4}, the Cauchy problem for the Wheeler-DeWitt equation
\begin{equation}\lae{3.63}
Hu=0
\end{equation}
can be uniquely solved in $E$ for given initial values on a Cauchy hypersurface, and the standard techniques of QFT can be applied to quantize the fields $u$ in \re{3.63} which represent gravitation. 

This procedure will be outlined in the sections below. Let us also mention that the constructed quantum field $\F_{\hat E}$, $\hat E$ will be a Cauchy hypersurface in $E$, will satisfy the Wheeler-DeWitt equation in a distributional sense indicating that the QFT approach is an accordance with the Hamiltonian constraint.

\section{The fibers are globally hyperbolic}\las{4}

DeWitt already analyzed the fibers in \cite{dewitt:qg}, though he did not look at them as fibers. Some of the ideas that we shall use in the proofs of the lemma and the theorem below can already be found in DeWitt's paper as we discovered after studying the literature.

\bl\lal{4.1}
Let $F$ be a fiber, then $F$ is connected and
\begin{equation}
\tau=\log\f
\end{equation}
is a time function satisfying
\begin{equation}\lae{4.2}  
\al_N\f^{-1}G^{ab}\tau_a\tau_b=-\frac n{4(n-1)}\al_N\f^{-1}.
\end{equation}
\el
\bp
$F$ is obviously connected, since $F$ is a convex cone in the vector space defined by the symmetric covariant tensors of order two.

To prove \re{4.2} we use the original coordinate representation $g_{ij}$ and conclude
\begin{equation}\lae{4.3}
\tau^{ij}=\pde\tau{g_{ij}}=\tfrac12 g^{ij},
\end{equation}
and hence
\begin{equation}
G_{ij,kl}\tau^{ij}\tau^{kl}=-\frac n{4(n-1)},
\end{equation}
where
\begin{equation}\lae{4.5}
G_{ij,kl}=\tfrac12\{g_{ik}g_{jl}+g_{il}g_{jk}\}-\tfrac1{n-1}g_{ij}g_{kl}
\end{equation}
is the inverse of $G^{ij,kl}$, hence the result.
\ep

\bt
Each fiber $F$ is globally hyperbolic, the hypersurface
\begin{equation}\lae{4.6} 
M=\{\f=1\}=\{\tau=0\}
\end{equation}
is a Cauchy hypersurface and in the corresponding Gaussian coordinate system $(\xi^a)$ the metric $\al^{-1}_N\f G_{ab}$ can be expressed as
\begin{equation}\lae{4.7}
ds^2=\frac {4(n-1)}n\al^{-1}_N\f\{-d\tau^2+G_{AB}d\xi^Ad\xi^B\},
\end{equation}
where
\begin{equation}
\tau=\xi^0\q\wed\q -\un<\tau<\un
\end{equation}
and $(\xi^A)$, $1\le A\le m$ are local coordinates for $M$. The metric $G_{AB}$ is also static, i.e., it does not depend on $\tau$.
\et
\bp
(i) Let $\tau$ be as in \rl{4.1}, then $\tau(F)=\R[]$ and in the conformal metric
\begin{equation}
\tilde G_{ab}=\al_N\f^{-1}\frac n{4(n-1)}(\al^{-1}_N\f G_{ab})
\end{equation}
$\tau_a$ is a unit gradient field in view of \re{4.2}.

(ii) The hypersurface $M$ in \re{4.6} is therefore spacelike and has at most countably many connected components.

Consider the flow
\begin{equation}
\begin{aligned}
\dot\xi&=-D\tau=-(\tilde G^{ab}\tau_b)\\
\xi(0,\zeta)&=\zeta,\q\zeta\in M.
\end{aligned}
\end{equation}
It will be convenient to express the flow in the original coordinate system, i.e.,
\begin{equation}
\begin{aligned}
\dot g_{ij}&=-\frac{4(n-1)}nG_{ij,kl}\tau^{kl},\\
g_{ij}(0,\zeta)&=\zeta=\bar g_{ij} ,
\end{aligned}
\end{equation}
where $G_{ij,kl}$ is the metric in \re{4.5}. The flow exists on a maximal time interval $J_\zeta$.

From \re{4.5} we obtain
\begin{equation}
\begin{aligned}
G_{ij,kl}\tau^{kl}&=\tfrac12 G_{ij,kl}g^{kl}\\
&=\tfrac12 g_{ij}(1-\frac n{n-1})=-\frac1{2(n-1)}g_{ij},
\end{aligned}
\end{equation}
hence
\begin{equation}\lae{4.13} 
\dot g_{ij}=\tfrac2ng_{ij}.
\end{equation}

Let $(\h^i)\in T^{1,0}_x(\so)$ be an arbitrary unit vector with respect to the metric $\chi_{ij}$, then
\begin{equation}
(g_{ij}\h^i\h^j)'=\tfrac2n g_{ij}\h^i\h^j
\end{equation}
leading to
\begin{equation}
g_{ij}\h^i\h^j=\bar g_{ij}\h^i\h^j e^{\frac2n t},
\end{equation}
thus the eigenvalues of $g_{ij}$ with respect to $\chi_{ij}$ are uniformly bounded from above and strictly bounded against zero when $\abs t\le \const$. Moreover,
\begin{equation}
\tau(g_{ij})=t
\end{equation}
from which we conclude
\begin{equation}
J_\zeta=\R[].
\end{equation}

If $M$ would be connected, then we would have proved that $F$ is product
\begin{equation}
F=\R[]\times M
\end{equation}
and that the metric would split as \re{4.7}. However, if $M$ had more then one connected component, then the corresponding cylinders defined by the flow would be disjoint and hence $F$ would not be connected.

(iii) Let $(\xi^a)$, $0\le a\le m$, be the corresponding Gaussian coordinate system such that
\begin{equation}
\xi^0=\tau=t
\end{equation}
and $(\xi^A)$, $1\le A\le m$, are local coordinates for $M$. Let $g_{ij}(\xi^a)$ be a local embedding in the new coordinate system, where the ambient metric should be the conformal metric up to a multiplicative constant, i.e., we consider
\begin{equation}
G^{ij,kl}=\tfrac12\{g^{ij}g^{kl}+g^{il}g^{jk}\}-g^{ij}g^{kl}
\end{equation}
to be the ambient metric such that
\begin{equation}
G_{ab}=G^{ij,kl}g_{ij,a}g_{kl,b}.
\end{equation}
The metric splits and we claim that
\begin{equation}
G_{AB}=G^{ij,kl}g_{ij,A}g_{kl,B}
\end{equation}
is stationary
\begin{equation}
\df{}tG_{AB}=0.
\end{equation}

To prove this equation we observe that the normal to $M(t)=\{\tau=t\}$ is a multiple of $g^{ij}$,  \cf\re{4.3}, hence
\begin{equation}
g^{ij}g_{ij,A}=0
\end{equation}
for $g_{ij}(t,\xi^A)$ is a local embedding of $M(t)$ from which we deduce
\begin{equation}\lae{4.25}
G_{AB}=\tfrac12\{g^{ik}g^{jl}+g^{il}g^{jk}\}g_{ij,A}g_{kl,B}.
\end{equation}
Differentiating this equation with respect to $t$ we infer, in view of \re{4.13},
\begin{equation}
\begin{aligned}
\df{}tG_{AB}&=-\tfrac2n\{g^{ik}g^{jl}+g^{il}g^{jk}\}g_{ij,A}g_{kl,B}\\
&\q+\tfrac2n\{g^{ik}g^{jl}+g^{il}g^{jk}\}g_{ij,A}g_{kl,B}\\
&=0
\end{aligned}
\end{equation}
where we also used
\begin{equation}
\dot g^{ij}=-\tfrac2ng^{ij}.
\end{equation}

(iv) Finally, we want to prove that $M=M(0)$ is a Cauchy hypersurface and hence $F$ globally hyperbolic, \cf \cite[Corollary 39, p. 422]{bn}. It suffices to prove this result for a conformal metric $G_{ab}$ where
\begin{equation}
d\bar s^2=-d\tau^2+G_{AB} d\xi^Ad\xi^B
\end{equation}
and $G_{AB}$ is stationary.

$G_{AB}$ is the metric of $M$. In case $n=3$ DeWitt proved in \cite[Remarks past equ. (5.15)]{dewitt:qg} that $M$ is a symmetric space and hence complete. DeWitt's proof in \cite[Appendix A]{dewitt:qg} remains valid for $n>3$. We shall only use the fact that $M$ is complete; in \rl{4.3b} below we shall give a second proof which does not rely on DeWitt's result.

Let $\ga(s)=(\ga^a(s))$, $s\in I$, be an inextendible future directed causal  curve in $F$ and assume that $\ga$ does not intersect $M$. We shall show that this will lead to a contradiction. It is also obvious that $\ga$ can meet $M$ at most once.

Assume that there exists $s_0\in I$ such that
\begin{equation}
\tau(\ga(s_0))<0
\end{equation}
and assume from now on that $s_0$ is the left endpoint of $I$. Since $\tau$ is continuous the whole curve $\ga$ must be contained in the past of $M$.

$\ga$ is causal, i.e.,
\begin{equation}
G_{AB}\dot\ga^A\dot\ga^B\le \abs{\dot\ga^0}^2
\end{equation}
and thus
\begin{equation}
\sqrt{G_{AB}\dot\ga^A\dot\ga^B}\le \dot\ga^0
\end{equation}
since $\ga$ is future directed. Let
\begin{equation}\lae{4.32} 
\tilde\ga=(\ga^A)
\end{equation}
be the projection of $\ga$ to $M$, then the length of $\tilde \ga$ is bounded
\begin{equation}\lae{4.33}
L(\tilde\ga)=\int_I\sqrt{G_{AB}\dot\ga^A\dot\ga^B}\le \int_I\dot\ga^0\le-\ga^0(s_0).
\end{equation}
Hence, $\tilde\ga$ stays in a compact set since $M$ is complete and the timelike coefficient is also bounded
\begin{equation}
\ga^0(s_0)\le \ga^0(s)<0\qq\A\, s\in I,
\end{equation}
which is a contradiction since $\ga$ should be inextendible but stays in a compact set of $F$.
\ep

\bl\lal{4.3b} 
The hypersurface $M=M(x)$ is a Cauchy hypersurface in $F(x)$.
\el
\bp
As in the proof above we consider an inextendible causal curve $\ga$ and look at the projection $\tilde\ga$ given in \re{4.32} which has finite length, \cf \re{4.33}. Then, it suffices to prove that $\tilde\ga$ stays in a compact subset of $M$.

Representing $\tilde\ga=\tilde\ga(s)$, $s\in I=[s_0,b)$, in the original coordinate system $(g_{ij})$
\begin{equation}
\tilde\ga=(g_{ij}(x,s))\equiv (g_{ij}(s))
\end{equation}
we use \re{4.25} to deduce
\begin{equation}
L(\tilde\ga)=\int_I\norm{\dot g_{ij}}\le -\ga^0(s_0),
\end{equation}
where
\begin{equation}
\norm{\dot g_{ij}}^2=g^{ik}g^{jl}\dot g_{ij}\dot g_{kl},
\end{equation}
from which we infer, in view of \cite[Lemma 14.2]{hamilton:ricci}, that the metrics $(g_{ij}(s))$ are all uniformly equivalent in $I$ and converge to a positive definite metric when $s\ra b$. Hence, the limit metric belongs to $M$ and $\tilde \ga$ stays in a compact subset of $M$.
\ep

\bt\lat{4.3}
Each fiber $F$ is a spacetime with a past crushing singularity, a big bang. It is past-timelike incomplete and the volume of the slices $M(\tau)$ converges to $0$ as $\tau\ra-\un$, or equivalently, when $\f\ra 0$.
\et
\bp
The mean curvature $\bar H$ of the slices $M(\tau)$ with respect to the past directed normal is
\begin{equation}
\bar H=-\frac m4\sqrt{\frac n{n-1}} \sqrt\al_N \f^{-1/2},
\end{equation}
as one easily checks. Moreover, let $\nu$ be their past directed normal vector field, then
\begin{equation}
 R_{ab}\nu^a\nu^b=0,
\end{equation}
where $(R_{ab})$ is the Ricci tensor of the ambient space, \cf \cite[equ. (A35) in Appendix A]{dewitt:qg}. Hence, Hawking's singularity theorem can be applied to conclude that $F$ is past-timelike incomplete. Furthermore, the sectional curvatures squared of the Riemann tensor tend to infinity if the singularity is approached, see \cite[Appendix A]{dewitt:qg}.
\ep

\section{Solving the Wheeler-DeWitt equation}\las{5}
In this section we want to solve the Wheeler-DeWitt equation \fre{3.59} in the bundle $E$. Let us first recall some well-known definitions and results for hyperbolic differential operators of second order in globally hyperbolic spacetimes.
\bd
Let $N$ be a smooth, globally hyperbolic spacetime. A linear differential operator $P$ of order two in $N$ is said to be \tit{normally} hyperbolic if it can be expressed in the form
\begin{equation}
P=-\D+b^\al D_\al+c,
\end{equation}
where $\D$ is the Laplace operator of the underlying metric, $(b^\al)$ a smooth vector field and $c$ a smooth function.
\ed

\bt
Let $N$ be globally hyperbolic, $M\su N$ a Cauchy hypersurface with future directed normal $\nu$, $P$ a normally hyperbolic operator, $u_0$, $u_1$ \resp $f$ functions in $C^\un_c(M,\K)$ \resp $C^\un_c(N,\K)$, then the Cauchy problem
\begin{equation}
\begin{aligned}
Pu&=f,\\
\fv uM&=u_0,\\
\fv{u_\al\nu^\al}M&=u_1,
\end{aligned}
\end{equation}
has a unique solution $u\in C^\un(N,\K)$ such that
\begin{equation}
\supp u\su J^N(K),
\end{equation}
where 
\begin{equation}
K=\supp u_0\,\uu\,\supp u_1\,\uu\,\supp f
\end{equation}
and
\begin{equation}
J^N(K)=J^N_+(K)\,\uu\,J^N_-(K),
\end{equation}
these are the points that can be reached by causal curves starting in $K$. Moreover, $u$ depends continuously on the data $(u_0,u_1,f)$ with corresponding estimates, namely, for any compact sets $K, K_1\su N$ and $K_0\su M$ and any $m\in\N$ there exists $m'\in\N$ and a constant $c=c(m, m', K,K_0,K_1)$ such that
\begin{equation}
\abs u_{m,K}\le c\{\abs{u_0}_{m',K_0}+\abs {u_1}_{m',K_0}+\abs f_{m',K_1}\},
\end{equation}
where $u$ is a solution of the Cauchy problem and $u_0$, $u_1$ and $f$ have support in the respective sets $K_0$ and $K_1$.
\et

A proof is given in \cite[Theorem 3.2.11, Theorem 3.2.12]{baer:book}.

\br
A corresponding result, when the data have no compact support, is also valid, \cf \cite[Corollary 5 on p. 78]{ginoux:fredenhagen}. In this case the estimates are of the form that for any compact $K\su N$ there is a compact $K'\su N$ such that for any $m\in\N$ there exists $m'\in\N$ and a constant $c$ such that
\begin{equation}\lae{5.7}
\abs u_{m,K}\le c\{\abs {u_0}_{m',K'\ii M}+\abs{u_1}_{m',K'\ii M}+\abs f_{m',K'}\}
\end{equation}
for all solutions $u$ of the Cauchy problem with smooth data $(u_0,u_1,f)$.
\er

The Hamilton operator in \fre{3.56} is certainly normally hyperbolic in each fiber, hence, given
\begin{equation}\lae{5.8}
u_0,u_1\in C^\un_c(\hat E,\K)
\end{equation}
and
\begin{equation}\lae{5.9}
f\in C^\un_c(E,\K)
\end{equation}
the Cauchy problem
\begin{equation}\lae{5.10}
\begin{aligned}
Hu&=f\\
\fv uM&=u_0\\
\fv{u_\al\nu^\al}M&=u_1
\end{aligned}
\end{equation}
has a unique solution
\begin{equation}
u=u(x,\xi(x))
\end{equation}
where $x\in \so$, $\xi^a(x)$ are local coordinates for the fibers $F(x)$ and $\hat E$ is the bundle with base space $\so$ and fibers $M(x)$, such that
\begin{equation}
u(x,\cdot)\in C^\un(F(x),\K).
\end{equation}
We shall  prove in the theorem below that the solutions are also smooth in $x$ such that $u\in C^\un(E,\K)$.
\bt
The solution $u$ of the Cauchy problem \re{5.10} with the data given in \re{5.8} and \re{5.9} belongs to the space $C^\un(E,\K)$ and satisfies
\begin{equation}
\supp u\su J^E(K)=\uuu_{x\in \so} J^{F(x)}(K(x)),
\end{equation}
where
\begin{equation}
K=\supp u_0\,\uu\,\supp u_1\,\uu\,\supp f
\end{equation}and
\begin{equation}
K(x)=K\ii \pi^{-1}(x),\qq x\in \so,
\end{equation}
and $\pi$ is the projection from $E$ to $\so$.
\et
\bp
We may work in a local trivialization $(\xi,U)$ of $E$
\begin{equation}
\xi:U\times (\Om\su \R[m+1])\ra \pi^{-1}(U)
\end{equation}
such that
\begin{equation}
\xi=\xi(x,\zeta),\qq \zeta\in\Om.
\end{equation}
Since the fibers are manifolds $\xi$ can be expressed in coordinates $\xi=(\xi^a(x,\zeta))$. A function  $u=u(\xi)$ can then  also be written in the form
\begin{equation}
u(x,\zeta)=u(\xi(x,\zeta)).
\end{equation}
The coordinates $\zeta=(\zeta^a)$ are then coordinates for the fibers $F(x)$, $x\in U$, such that the differential operator $P$ has the form
\begin{equation}
Pu=-G^{ab}u_{ab}+b^au_a+cu,
\end{equation}
where the derivatives are partial derivatives with respect to $\zeta$ and the coefficients and $u$ depend on $(x,\zeta)$. We shall prove that $u(x,\zeta)$ is smooth in $x$ by using the difference quotient method. For simplicity we use the symbol $\xi$ instead of $\zeta$ to denote the coordinates.

(i) We shall first prove that $u$ is Lipschitz continuous in $x$. Let $x=(x^i)$, fix $1\le k\le n$ and let $h\ne 0$ be a real number. By a slight abuse of notation we define the $n$-tuple
\begin{equation}
h=(0,\dots,0,h,0,\dots,0),
\end{equation}
where only the $k$-th component is different from $0$. 

Let
\begin{equation}\lae{5.20}
v=h^{-1}(u(x+h,\cdot)-u(x,\cdot)),
\end{equation}
then $v$ solves the differential equation
\begin{equation}\lae{5.21}
Pv=f_h,
\end{equation}
where $f_h(x,\cdot)\in C^\un(F(x),\K)$ such that for any given compact $K\su E$ and $p\in\N$
\begin{equation}\lae{5.22}
\abs{f_h}_{p,K\ii F(x)}\le c=c(p,K)\qq\A\, \abs h<\e_0
\end{equation}
uniformly in $x$, with initial values
\begin{equation}
h^{-1}(u_0(x+h,\cdot)-u(x,\cdot))
\end{equation}
and
\begin{equation}
h^{-1}(u_1(x+h,\cdot)-u_1(x,\cdot)).
\end{equation}

To verify \re{5.21}, \re{5.22} we only consider the main part of the differential operator
\begin{equation}
G^{ab}(x+h)u_{ab}(x+h)-G^{ab}(x)u_{ab}(x),
\end{equation}
where $u_{ab}$ are partial derivatives and use the algebraic identity
\begin{equation}
\begin{aligned}
&G^{ab}(x+h)u_{ab}(x+h)-G^{ab}(x)u_{ab}(x)\\
&=G^{ab}(x)\{u(x+h)-u(x)\}_{ab}+\{G^{ab}(x+h)-G^{ab}(x)\}u_{ab}(x+h).
\end{aligned}
\end{equation}
Similar expressions occur when we look at the lower order terms. Hence \re{5.21} and \re{5.22} are proved.

Then we apply the a priori estimates \re{5.7} to conclude
\begin{equation}
\abs v_{p,K\ii F(x)}\le c(p,K)
\end{equation}
for any compact subset of $K\su E$  and any $p\in\N$ independent of $h$ and $x\in U$. Therefore $u$ and its derivatives with respect to $\xi$ are Lipschitz continuous in $x$.

(ii) Next we shall prove that $u$ is of class $C^1$ in $x$. Differentiate the equation
\begin{equation}
Pu=f
\end{equation}
with respect $x^k$, $1\le k\le n$, and pretend that $u$ can be differentiated, then we obtain
\begin{equation}
Pu_k=\tilde f,
\end{equation}
where
\begin{equation}
u_k=\pde u{x^k}
\end{equation}
and $\tilde f(x,\cdot)\in C^\un(F(x),\K)$ and Lipschitz continuous in $x$.

Let $w$ be a solution of the Cauchy problem
\begin{equation}
Pw=\tilde f
\end{equation}
with initial values
\begin{equation}
\pde{u_0}{x^k}\q\wed\q \pde{u_1}{x^k},
\end{equation}
$w=w(x,\xi)$ is of class $C^\un$ in $\xi$ and Lipschitz continuous in $x$ because of the arguments in (i). When we look at the difference
\begin{equation}
P(v-w)=f_h-\tilde f,
\end{equation}
where $v$ is defined in \re{5.20}, we deduce that the right-hand side converges to zero in the limit $h\ra0$.

The same result is also valid for the difference of the initial values, hence
\begin{equation}
\pde u{x^k}=w
\end{equation}
and $u$ is of class $C^1$ in $x$ as well as its derivatives with respect to $\xi$.

(iii) We can now prove by induction that $u$ is in $C^\un(E,\K)$.
\ep

\bc
Let $P$ be as in the preceding theorem. For $(x,\xi)\in E$ let $F_\pm$ be the fundamental solutions with respect to $P$ with past-compact \resp future-compact support at $(x,\xi)$ and let $G_\pm$ be the advanced \resp retarded Green's operators of $P$. Then, for any $u\in C^\un_c(E,\K)$, the mappings
\begin{equation}\lae{5.35}
(x,\xi)\ra F_\pm(x,\xi)[u]
\end{equation}
are smooth and
\begin{equation}\lae{5.36}
G_\pm(u)\in C^\un(E,\K).
\end{equation}
\ec
\bp
\cq{\re{5.35}}\q We only prove the claim for $F_+$. There exists a a  function $\chi(u)$, which is given as the solution of the Cauchy problem
\begin{equation}
P^*\chi(u)=u
\end{equation}
with vanishing initial values on a Cauchy hypersurface, where $P^*$ is the adjoint of $P$---we do not assume $P$ to be formally self-adjoint---such that
\begin{equation}
F_+(x,\xi)[u]=\chi(u),
\end{equation}
\cf \cite[equ. (3.10) on p. 79]{ginoux:fredenhagen}. Since $P^*$ is also normally hyperbolic and smooth, $\chi(u)\in C^\un(E,\K))$ because of the preceding theorem and the claim is proved.

\cvm
\cq{\re{5.36}}\q The Green's operators are defined by
\begin{equation}
(G_\mp u)(x,\xi)=F_\pm(x,\xi)[u]\qq\A\, (x,\xi)\in E,
\end{equation}
hence the result because of \re{5.35}.
\ep
\section{Applying QFT techniques to the Wheeler-DeWitt equation}\las{6}
The main difference between the standard QFT setting and the present situation is that we consider a bundle where the fibers are globally hyperbolic spacetimes and the hyperbolic operator is acting in the fibers. However, because of the results in \rs{5}, we can solve the Cauchy problem for the Hamiltonian operator $H$ in the bundle $E$, hence there exist the advanced and retarded Green distributions $G_+$ and $G_-$ for $H$ such that
\begin{equation}
G_\pm:C^\un_c(E,\K)\ra C^\un(E,\K)
\end{equation}
\begin{equation}\lae{6.2}
H\circ G_\pm=G_\pm\circ \fv H{C^\un_c(E,\K)}=\id_{C^\un_c(E,\K)}
\end{equation}
\begin{equation}
\supp \,(G_\pm u)\su J^E_+(\supp u)\qq\A\, u\in C^\un_c(E,\K)
\end{equation}
and
\begin{equation}\lae{6.4}
\supp\,(G_- u)\su J^E_-(\supp u)\qq\A\, u\in C^\un_c(E,\K).
\end{equation}

$H$ is formally self-adjoint
\begin{equation}
\begin{aligned}
\spd{Hu}v_E&=\int_\so\int_{F}Hu\bar v\\
&=\int_\so\int_{F}u\overline{Hv}=\spd u{Hv}_E,
\end{aligned}
\end{equation}
where the volume element in $F$ is given by the Lorentzian metric and that in $\so$ is defined by the previously used metric $\chi$.

There are two ways to construct a quantum field given a formally self-adjoint normally hyperbolic operator in a globally hyperbolic spacetime  or, in our case, in the bundle $E$. One possibility is to define a symplectic vector space
\begin{equation}
V=C^\un_c(E,\K)/N(G),
\end{equation}
where $G$ is the Green's distribution
\begin{equation}
G=G_+-G_-.
\end{equation}
Since
\begin{equation}
G^*=-G
\end{equation}
the bilinear form
\begin{equation}
\int_\so\int_F\spd{Gu}{v}\qq u,v\in V
\end{equation}
is skew-symmetric,  non-degenerate by definition and hence symplectic, and then there is a canonical way to construct a $C^*$-algebra.

The second method is to use a Cauchy hypersurface to define a quantum field in Fock space. We shall show that this is also possible in our case.

First let us prove the following lemma:
\bl
For all $u,v\in C^\un_c(\hat E,\K)$ there holds
\begin{equation}
\begin{aligned}
\int_\so\int_F\spd u{Gv}=\int_\so\int_M\{\spd{D_\nu(Gu)}{Gv}-\spd{Gu}{D_\nu(Gv)}\},
\end{aligned}
\end{equation}
where $\nu$ is the future normal to $M$ and the scalar product is the standard scalar product in $\K$.
\el 
\bp
Let $F=F(x)$ be an arbitrary fiber, then
\begin{equation}
\int_F\spd u{Gv}=\int_{F_+}\spd u{Gv}+\int_{F_-}\spd u{Gv},
\end{equation}
where
\begin{equation}
F_+=\{\xi^0>0\}\q\wed\q F_-=\{\xi^0<0\}.
\end{equation}
Now, in $F_+$ we have
\begin{equation}
HG_-u=u
\end{equation}
and
\begin{equation}
HGv=0,
\end{equation}
in view of \re{6.2}. Moreover, because of \re{6.4}, we have 
\begin{equation}
\supp\,(G_-u)\ii\bar F_+\q \text{is compact},
\end{equation}
since
\begin{equation}\lae{6.16}
\supp\,(G_-u)\ii M\q\text{is compact}.
\end{equation}
Hence, we obtain by partial integration
\begin{equation}
\int_{F_+}\spd{HG_-u}{Gv}=-\int_M\spd{D_\nu G_-u}{Gv}+\int_M\spd{G_-u}{D_\nu Gv}.
\end{equation}
A similar argument applies in $F_-$ by looking at
\begin{equation}
HG_+u=u
\end{equation}
leading to
\begin{equation}
\int_{F_-}\spd{HG_+u}{Gv}=\int_M\spd{D_\nu G_+u}{Gv}-\int_M\spd{G_+u}{D_\nu Gv}.
\end{equation}
Adding these two relations implies the result in the fiber.
\ep
We now define the Hilbert space $H_{\hat E}$ which is used to construct the symmetric Fock space, namely, we set
\begin{equation}
H_{\hat E}=L^2(\hat E,\Cc)
\end{equation}
with the standard scalar product
\begin{equation}
\spd uv_{\hat E}=\int_\so\int_M\spd uv.
\end{equation}

We denote the symmetric Fock space by $\mc F(H_{\hat E})$. Let $\Theta$ be the corresponding Segal field. Since $G^*=-G$ we deduce from  \re{6.16} and from a similar result for $G_+$
\begin{equation}
\fv {G^*u}{\hat E}\in C^\un_c(\hat E,\R[])\su H_{\hat E}.
\end{equation}
We can therefore define
\begin{equation}\lae{6.23}
\F_{\hat E}(u)=\Theta(i\fv {(G^*u)}{\hat E}-D_\nu\fv{(G^*u)}{\hat E}).
\end{equation}
From the proof of \cite[Lemma 4.6.8]{baer:book} we conclude that the right-hand side of  \re{6.23} is an essentially self-adjoint operator in $\mc F(H_{\hat E})$. We therefore call the map $\F_{\hat E}$ from $C^\un_c(E,\R[])$ to the set of self-adjoint operators in $\mc F(H_{\hat E})$ a quantum field for $H$ defined by $\hat E$.

\bl
The quantum field $\F_{\hat E}$ satisfies the equation
\begin{equation}
H\F_{\hat E}=0
\end{equation}
in the distributional sense, i.e.,
\begin{equation}
\spd{H\F_{\hat E}}u=\spd{\F_{\hat E}}{Hu}=\F_{\hat E}(Hu)=0\q\A\, u\in C^\un_c(E,\R[]).
\end{equation}
\el
\bp
In view of \re{6.2} there holds
\begin{equation}
G^*Hu=0\qq\A\, u\in C^\un_c(E,\R[]).
\end{equation}
\ep

These remarks should suffice to indicate that the techniques of QFT can be applied in the present case. For further details we refer to \cite[Chapter 4.7]{baer:book}.

\section{Gravity interacting with a scalar field}
In this section we demonstrate that other fields can be added to the functional and that the general procedure, define a fiber bundle where the fibers are globally hyperbolic spacetimes, can naturally be applied---at least in the case of a scalar field.

Thus, let $M_1$ be a complete Riemannian  space of dimension $n_1$ and let
\begin{equation}
y:N\ra M_1
\end{equation}
be a scalar field map; in local coordinates we have
\begin{equation}
y=(y^A(x^\al)).
\end{equation}
$N$ is the Lorentzian manifold we looked at in \rs{3}.

We shall consider the Lagrangian
\begin{equation}
L_1=-\tfrac12 \bar g^{\al\bet}y^A_\al y^B_\bet g_{AB}-V(y),
\end{equation}
where $(g_{AB})$ is the fixed metric in $M_1$ and $V(y)$ a smooth potential.

After introducing the time function $x^0$ and the Cauchy hypersurface $\so$ $L_1$ can be written in the form
\begin{equation}
L_1=\tfrac12 w^{-2}g_{AB}\dot y^A\dot y^B-\tfrac12 g^{ij}y^A_i y^B_j g_{AB} -V(y).
\end{equation}
We still have to multiply this Lagrangian with $w\sqrt g$. Introducing the function $\f$ as in \fre{3.42}, we obtain
\begin{equation}
\begin{aligned}
L_1=\tfrac12 w^{-1}g_{AB}\dot y^A\dot y^B\f-\tfrac12 g^{ij}y^A_i y^B_j g_{ab}w\f-V(y)w\f,
\end{aligned}
\end{equation}
where we omit $\sqrt\chi$ as before.

Similar as in \rs{3} we regard the mapping $y$ as a section in a bundle with base manifold $\so$ and fibers $M_1$. Since the metrics $g_{ij}$ are also looked at as the section of a fiber bundle, we define the bundle $E$ as a fiber bundle with base space $\so$ and fibers
\begin{equation}
F(x)\times M_1,
\end{equation}
where the fibers are a metric product.

Defining
\begin{equation}
p_A=\pde{L_1}{\dot y^A}
\end{equation}
we obtain
\begin{equation}
\begin{aligned}
p_A\pde{L_1}{\dot y^A}-L_1&=\tfrac12g_{AB}(\dot y^Aw^{-1}\f)(\dot y^Bw^{-1}\f)w\f^{-1}\\
&\q+\tfrac12 g^{ij}y^A_iy^B_j g_{AB} w\f+V(y)w\f.
\end{aligned}
\end{equation}
Hence the combined Hamiltonian constraint has the form, compare with \fre{3.52},
\begin{equation}
\begin{aligned}
H&=\al_N\f^{-1}G^{ab}\pi_a\pi_b+\tfrac12\f^{-1}g^{AB}p_Ap_B\\
&\q+\tfrac12 g^{ij}y^A_iy^B_jg_{AB}\f-\al^{-1}_N(R-2\Lam)\f+V\f=0.
\end{aligned}
\end{equation}

The common product metric looks like
\begin{equation}\lae{7.10}
G=\f
\begin{pmatrix}
\al_N^{-1}G_{ab}&0\\[\cma]
0&2g_{AB}
\end{pmatrix}.
\end{equation}
After canonical quantization the Hamiltonian operator has the form
\begin{equation}
H=-\D-\al^{-1}_N(R-2\Lam)\f+\tfrac12g^{ij}y^A_iy^B_jg_{AB}\f+V\f
\end{equation}
and the Wheeler-DeWitt equation is
\begin{equation}
Hu=0.
\end{equation}
The term
\begin{equation}
g^{ij}y^A_iy^B_jg_{AB}
\end{equation}
is well defined since
\begin{equation}
(\xi^a(x),y^A(x))\equiv(g_{ij},y^A(x))
\end{equation}
is a smooth section in the bundle $E$.

The fibers
\begin{equation}
F\times M_1,
\end{equation}
 equipped with the product metric \re{7.10}, are globally hyperbolic spacetimes; the former Cauchy hypersurface $M$ has to be replaced by
 \begin{equation}
M\times M_1
\end{equation}
and the Gaussian coordinate system associated with the new Cauchy hypersurface is
\begin{equation}
(\xi^a,y^A).
\end{equation}

The proof that the fibers are globally hyperbolic is identical to the one given in \rs{4}. The previous considerations in \rs{5} and \rs{6} are therefore  valid in the present setting leading to a unified quantum theory. 

\nocite{thiemann:book,kiefer:book}
\bibliographystyle{hamsplain}

\providecommand{\bysame}{\leavevmode\hbox to3em{\hrulefill}\thinspace}
\providecommand{\href}[2]{#2}



\end{document}